\documentclass[runningheads,a4paper,11pt]{llncs}
\usepackage[DIV10,twoside=semi]{typearea}

\usepackage{amsmath, amssymb}
\usepackage{url, array, calc}
\usepackage[mathscr]{eucal}
\usepackage{cite}

\def\cC{\mathcal{C}} \def\cG{\mathcal{G}} \def\cL{\mathcal{L}}

\newcommand\abs[1]{\left\lvert#1\right\rvert}
\newcommand\ico{\mathit{ico}}
\newcommand\icz{\mathit{icz}}

\newcounter{rules}

\newcounter{saverules}

\newenvironment{rules}%
  {\begin{list}{}{%
    \usecounter{rules}
    \setcounter{rules}{\value{saverules}}%
    \setlength{\labelwidth}{40pt}%
    \setlength{\leftmargin}{\labelwidth+\labelsep}%
  }}%
  {\setcounter{saverules}{\value{rules}}\end{list}}

\begin{document}

\title{Towards a Data Reduction for the \\ Minimum Flip Supertree Problem}

\author{Sebastian B\"ocker}

\institute{Lehrstuhl f{\"u}r Bioinformatik
  \\ Friedrich-Schiller-Universit{\"a}t Jena \\ Ernst-Abbe-Platz 2 \\ 07743
  Jena, Germany \\ \email{sebastian.boecker@uni-jena.de}}

\maketitle


\begin{abstract}
  In computational phylogenetics, the problem of constructing a supertree of
  a given set of rooted input trees can be formalized in different ways, to
  cope with contradictory information in the input.  We consider the Minimum
  Flip Supertree problem, where the input trees are transformed into a
  0/1/?-matrix, such that each row represents a taxon, and each column
  represents an inner node of one of the input trees.  Our goal is to find a
  perfect phylogeny for the input matrix requiring a minimum number of
  0/1-flips, that is, corrections of 0/1-entries in the matrix.  The problem
  is known to be NP-complete.

  Here, we present a parameterized data reduction with polynomial running
  time.  The data reduction guarantees that the reduced instance has a
  solution if and only if the original instance has a solution.  We then make
  our data reduction parameter-independent by using upper bounds.  This
  allows us to preprocess an instance, and to solve the reduced instance with
  an arbitrary method.  Different from an existing data reduction for the
  consensus tree problem, our reduction allows us to draw conclusions about
  certain entries in the matrix.  We have implemented and evaluated our data
  reduction.  Unfortunately, we find that the Minimum Flip Supertree problem
  is also hard in practice: The amount of information that can be derived
  during data reduction diminishes as instances get more ``complicated'', and
  running times for ``complicated'' instances quickly become prohibitive.
  Still, our method offers another route of attack for this relevant
  phylogenetic problem.
\end{abstract}


\section{Introduction}

When studying the relationship and ancestry of current organisms, discovered
relations are usually represented as phylogenetic trees, that is, rooted
trees where each leaf corresponds to a group of organisms, called
\emph{taxon}, and inner vertices represent hypothetical last common ancestors
(or latest common ancestor) of the organisms located at the leaves of its
subtree.

Supertree methods assemble phylogenetic trees with non-identical but
overlapping taxon sets, into a larger supertree that contains all taxa of
every input tree and describes the evolutionary relationship of these
taxa~\cite{bininda-emonds04phylogenetic}.  Constructing a supertree is easy
if no contradictory information is encoded in the input
trees~\cite{ng96reconstruction, bryant95extension}.  The major problem of
supertree methods is dealing with incompatible data in a reasonable way,
where it should be understood that incompatible input trees are the rule
rather than the exception in phylogenetic supertree analysis.

Matrix representation (MR) supertree methods encode inner vertices of all
input trees as partial binary characters in a matrix, which is then analyzed
using an optimization or agreement criterion to yield the supertree.  In
1992, Baum~\cite{baum92combining} and Ragan~\cite{ragan92phylogenetic}
independently proposed the matrix representation with parsimony (MRP) method
as the first matrix representation method, that performs a maximum parsimony
analysis on a matrix representation of the input trees.  MRP is by far the
most widely used supertree method today, and constructed supertrees are of
comparatively high quality.  The Maximum Parsimony problem is
NP-complete~\cite{day86computational-a}, and so is the MRP problem.

The matrix representation with flipping (MRF) supertree method also uses a
matrix representation of the rooted input trees, with matrix entries `0',
`1', and `?'~\cite{chen06minimum-flip}.  Utilizing the parsimony principle,
MRF seeks the minimum number of ``flips'' $0 \to 1$ or $1 \to 0$ in the input
matrix that make the resulting matrix consistent with a phylogenetic tree,
where `?'-entries can be resolved arbitrarily.  Evaluations by Chen
\emph{et~al.}~\cite{chen06improved-heuristics} indicate that MRF is on par
with the ``gold standard'' MRP, and superior to other approaches for
supertree construction.  Most supertree methods take rooted trees as input,
and so does MRF; but this is not a problem in practice, as in practically all
relevant cases, input trees can be rooted by an outgroup.

If all input trees share the same set of taxa, the supertree is called a
consensus tree~\cite{adams72consensus}.  As for supertrees, we can encode the
input trees in a matrix, here with matrix entries `0' and `1'.  In case there
exist no conflicts between the input trees, we can construct the
corresponding \emph{perfect phylogeny} in $\Theta(mn)$ time for $n$ taxa and
$m$ characters~\cite{gusfield91efficient}.  To deal with incompatible input
trees, the MRF consensus tree problem again seeks the minimum number of flips
in the input matrix to reach a perfect phylogeny.  This problem is
NP-hard~\cite{chen06minimum-flip}, but there have been some recent
algorithmic results: The problem can be approximated with approximation ratio
$2d$ where $d$ is the maximum number of ones in a
column~\cite{chen06minimum-flip}.  This approximation ratio is obviously
prohibitive in practice, but no constant factor approximation is known.  On
the parameterized side, let $k$ denote the number of flips required to
correct the input matrix: Komusiewicz \emph{et~al.}~\cite{komusiewicz08cubic}
give a problem kernel with $O(k^3)$ vertices for the MRF consensus tree
problem, and B\"ocker \emph{et~al.}~\cite{boecker08improved} present a
$O(4.83^k + poly(m,n))$ search tree algorithm.

For the more general MRF supertree problem, there has been less progress:
Clearly, the MRF supertree decision problem is NP-complete, as it generalizes
the MRF consensus tree problem, and we can check in polynomial time if a
given binary matrix $M^*$ is a perfect phylogeny and has distance at most $k$
to our input matrix.  We can test whether an MRF supertree instance admits a
perfect phylogeny without flipping in time $\tilde{O}
(mn)$~\cite{peer04incomplete}.  There exist no approximation algorithms or
parameterized algorithms in the literature.  Chen
\emph{et~al.}~\cite{chen06improved-heuristics} present a heuristic for MRF
supertrees based on branch swapping, and Chimani
\emph{et~al.}~\cite{chimani10exact} introduce an Integer Linear Program (ILP)
to find exact solutions.  Recently, B\"ocker
\emph{et~al.}~\cite{brinkmeyer11flipcut} presented a heuristic top-down
algorithm based on the MRF intuition, namely the \textsc{FlipCut} supertree
method, which is both swift and accurate in practice.

\paragraph*{Our contributions.}

Here, we present a set of reduction rules that can be applied to an arbitrary
instance of the MRF supertree problem, requiring polynomial running time.
Our data reduction is parameterized, in the sense that we assume a maximal
number of flips $k$ to be given.  The data reduction guarantees that the
reduced instance has a solution if and only if the original instance has a
solution.  We then show how to make the reduction parameter-independent, by
using upper and lower bounds. This allows us to preprocess an instance, and
to solve the reduced instance with any method, be it an ILP, a search tree
algorithm, or a heuristic.  Different from~\cite{komusiewicz08cubic}, our
data reduction allows us to draw conclusions about certain entries in the
input matrix, whereas the data reduction for MRF consensus trees
in~\cite{komusiewicz08cubic} only removes certain characters and taxa from
the input.

We have implemented and evaluated our data reduction on a set of MRF
supertree instances from~\cite{chen06improved-heuristics}.  Unfortunately, we
find that running times become prohibitive when instances become large, or
contain many~`?'.  This agrees with findings in~\cite{chimani10exact}, where
``complicated'' instances could not be processed by the ILP in reasonable
running time.  Still and all, we believe that the data reduction presented
here, can be an important step towards both exact methods and improved
heuristics for the MRF supertree problem.


\section{Preliminaries} \label{sec:preliminaries}

Let $n$ be the number of taxa and $m$ be the number of characters or
features.  For brevity, we assume that our set of characters equals
$\{1,\dots,m\}$, and that our set of taxa equals $\{1,\dots,n\}$.  Each taxon
$t$ can possess or not possess each character $v$, encoded in a \emph{binary}
$n \times m$ matrix $M$, where columns of $M$ correspond to characters and
rows correspond to taxa.  For the moment, we do not allow `?' to appear in
the input matrix.  Under the classical \emph{perfect phylogeny}
model~\cite{wilson65consistency}, we assume that there exists an ancestral
species that possesses none of the characters, corresponding to a row of
zeros.  We further assume that each transition from `0' to `1' happens at
most once in the tree: An invented character never disappears and is never
invented twice.  We say that $M$ \emph{admits a perfect phylogeny} if there
is a rooted tree with $n$ leaves corresponding to the $n$ taxa, where for
each character $u$, there is an inner node $w$ of the tree such that
$M[t,u]=1$ holds if and only if taxon $t$ is a leaf of the subtree below~$u$,
for all~$t$.

Given an arbitrary binary matrix $M$, we may ask whether $M$ admits a perfect
phylogeny.  Gusfield~\cite{gusfield91efficient} shows how to test $M$ and, if
possible, construct the corresponding phylogenetic tree in time $\Theta(mn)$.
There exist several characterizations for such
matrices~\cite{peer04incomplete}, of which we only mention two here.  Let
$I_M(v) := \{t: M[t,v] = 1\}$ be the set of `1'-indices in column~$v$.
Matrices that admit a perfect phylogeny, can be characterized via the
pairwise \emph{compatibility} of all column pairs $u,v$: That is, $I_M(u)
\subseteq I_M(v)$ or $I_M(v) \subseteq I_M(u)$ or $I_M(u) \cap I_M(v) =
\emptyset$ must hold.  Characters that do not satisfy this condition are said
to be \emph{in conflict}.  We can also characterize such matrices via local
conflicts: Let $\cG(M)$ be the bipartite graph on character vertices $v \in
\{1,\dots,m\}$ and taxa vertices $t \in \{1,\dots,n\}$, such that an edge
$(v,t)$ exists if and only if $M[t,v]=1$.  Now, $M$ admits a perfect
phylogeny if and only if the graph $\cG(M)$ is \emph{M-free}, that is, it
does not contain an induced path of length four starting from and ending in
different taxa vertices~\cite{chen06minimum-flip}.

We consider two variants of Matrix Representation with Flipping problems,
namely the \textsc{Minimum Flip Consensus Tree} (MFCT) and the
\textsc{Minimum Flip Supertree} (MFST) problem.  For the MFCT problem,
consider a set of binary rooted trees on the same set of $n$ taxa.  We encode
the input trees in a binary matrix $M$, where each column corresponds to an
inner node in one of the trees, and an entry `1' indicates that the
corresponding taxon is a leaf of the subtree rooted in the inner node.  We
ask for the minimum number of modifications (``flips'') to $M$ such that the
resulting matrix admits a perfect phylogeny.  We refer to this number of
flips as the \emph{cost} of the instance.

The more general MFST problem arises when the input trees have overlapping
but not necessarily identical taxa sets.  In this case, for characters
belonging to a particular input tree, the state (`0'~or~`1') of some taxa is
not known as they are not part of the input tree, and represented by a
question mark (`?').  We ask for a perfect phylogeny matrix $M^*$ such that
the number of entries where one matrix contains a `0' and the other matrix a
`1', is minimal.  This is the number of flips required to correct the input
matrix $M$, whereas `?'-entries can be resolved arbitrarily.
Note that a perfect phylogeny matrix must not contain `?' entries.  Both for
MFST and MFCT, we usually have $n \ll m$.

Throughout this paper, we assume that the input matrix $M$ does not contain
any all-zero columns: If the matrix would contain such columns, we could
simply remove them.  We infer that any optimal solution does not contain an
all-zero column: Otherwise, we could leave one of the entries in the flipped
matrix $M^*$ in its original state `1', thereby constructing a matrix that is
also a perfect phylogeny but requires less flipping.  This follows because a
character that is exhibited by a single taxon, cannot be in conflict with any
other character.  We make use of the fact throughout this paper without
explicitly referring to it.


\section{The inclusion graph}

Given an instance $M \in \{0,1\}^{n \times m}$ of the \textsc{Minimum Flip
  Consensus Tree} problem, we say that two characters $u$ and $v$ are
\emph{in conflict} if $I_M(u) \cap I_M(v) \ne \emptyset$ but $I_M(u)
\not\subseteq I_M(v)$ and $I_M(v) \not\subseteq I_M(u)$.  We define the
\emph{inclusion graph} $G = (V,E)$ as follows: This graph has vertex set $V
:= \{1,\dots,m\}$, being the characters of matrix~$M$.  Two vertices $u,v \in
V$ can be connected via a directed edge $(u,v)$, or by an undirected edge $uv
= \{u,v\}$.  An \emph{inclusion edge} $(u,v)$ from $u \in V$ to $v \in V$ is
present if $I_M(u) \subseteq I_M(v)$.  A \emph{disjoint edge} $uv$ connecting
$u,v \in V$ is present if $I_M(u) \cap I_M(v) = \emptyset$.  Any two vertices
$u,v$ are connected by either no edge in case $u,v$ are in conflict; by a
single edge $(u,v)$, $(v,u)$, or $uv$; or, by two inclusion edges $(u,v)$ and
$(v,u)$ at the same time.

If two vertices $u,v \in V$ are connected by both directed edges $(u,v)$ and
$(v,u)$, then $u$ and $v$ have the same neighborhood $I_M(u) = I_M(v)$.  In
this case, there exists an optimal solution such that $u$ and $v$ also have
the same neighborhood~\cite{komusiewicz08cubic}.  There may also exist
optimal solutions such that $u$ and $v$ have different neighborhoods, but
this case is somewhat pathogenic and will rarely appear in practice.  In view
of this, we can immediately \emph{merge} $u$ and~$v$.  In order to merge
nodes in the inclusion graph, we assume that each character vertex $v \in V$
and each column of the matrix $M$ has a weight assigned to it, representing
its multiplicity.  For readability, we omit these simple details in the
following.  Now, we may assume that any two vertices $u,v \in V$ are
connected by at most one edge.

We know that $M$ admits a perfect phylogeny if and only if there exist no two
vertices $u,v \in V$ that are in conflict.  This is the case if any two
vertices in the inclusion graph are connected by (at~least) one edge.  When
resolving conflicts in the matrix~$M$, this will lead to induced changes in
the inclusion graph.  This allows us to reformulate our problem: We search
for the minimum number of changes in~$M$, such that any two vertices in the
inclusion graph are connected by (at~least) one edge.

If $M$ admits a perfect phylogeny, then the resulting graph is
\emph{transitive}: from $(u,v) \in E$ and $(v,w) \in E$ we infer $(u,w) \in
E$.  But we can derive a similar deduction rule for disjoint edges: from
$(u,v) \in E$ and $vw \in E$ we infer $uw \in E$.  We say that an inclusion
graph is \emph{tree-ish} if it satisfies these deduction rules for all
vertices.

In applications, it is of no avail to actually compute the inclusion graph of
a matrix~$M$, as we can compute on the fly whether an edge is present or not,
using~$M$.  Still, the inclusion graph is useful in applications: During data
reduction,
we sometimes learn that, say, $I_M(u) \subseteq I_M(v)$ must hold for the
optimal solution.  In this case, we set the respective edge to ``permanent''.
More often, we will learn from the data that, say, $I_M(u) \subseteq I_M(v)$
cannot hold for the optimal solution.  In this case, we set the respective
edge to ``forbidden''.  Note that forbidden edges may co-exist in parallel
for one vertex pair $u,v$.  But in case two out of the three edges $(u,v)$,
$(v,u)$, and $uv$ are set to forbidden, we can immediately set the remaining
edge to permanent.

The inclusion graph, in turn, allows us to draw conclusions about entries
in~$M$: If there is a permanent edge $(u,v)$ in the inclusion graph, and we
decide to change or keep an entry $M[t,u] = 1$ in our input matrix, this
forces us to also set $M[t,v] = 1$.  Similarly, if we decide to change or
keep an entry $M[t,v] = 0$ then the edge $(u,v)$ in the inclusion graph also
forces us to set $M[t,u] = 0$.  We will formalize these observations in the
next section.

Note that we can define a similar inclusion graph for an instance $M \in
\{0,1,?\}^{n \times m}$ of the \textsc{Minimum Flip Supertree} problem.
Here, $M$ admitting a perfect phylogeny $M^*$ does not imply that every two
vertices in the inclusion graph are connected by an edge: For example, an
input matrix containing solely `?' results in a inclusion graph without
edges.  But our other reasoning introduced above, remains valid.


\section{Parameterized data reduction} \label{sec:data-reduction}

We now describe data reduction rules for the \textsc{Minimum Flip Supertree}
problem.  Here, entries in the matrix $M$ can be `?', and we have to assure
that such entries are chosen ``conservatively'': To this end, we define
$I^*_M(v) := \{t: M[t,v] \in \{1,?\}\}$.

We take a \emph{parameterized} view of the problem: We assume that we are
given an integer $k$, and we want to know if there exists a solution for
input matrix $M$ with cost at most~$k$.  This will allow us to set certain
edges of the inclusion graph to forbidden or permanent, and also to
permanently set certain entries in the matrix $M$, which may include
resolving `?'-entries or even flipping entries in the matrix.  We will see in
Sec.~\ref{sec:bounds} how these rules can be applied during preprocessing.

For $u, v \in V$ we set $N(u-v) := I_M(u) \setminus I^*_M(v)$ and $N(u+v) :=
I_M(u) \cap I_M(v)$.  Recall that $(u,v)$ being present in the inclusion
graph of an optimal solution $M^* \in \{0,1\}^{n \times m}$, implies that
$I_{M^*}(u) \subseteq I_{M^*}(v)$ must hold.  Similarly, $uv$ being present
implies that $I_{M^*}(u) \cap I_{M^*}(v) = \emptyset$.  As we assume that the
distance between $M^*$ and $M$ is at most $k$ flips, we can easily deduce two
simple reduction rules:
\begin{rules}
\item \label{rule:start} \label{rule:param-edges-start} If $\abs{N(u-v)} > k$
  then set $(u,v)$ to forbidden.

\item \label{rule:param-edges-end} If $\abs{N(u+v)} > k$ then set $uv$ to
  forbidden.
\end{rules}
Note that the first rule is two-sided, as edges $(u,v)$ are directed.  In
case two of the three possible edges $(u,v)$, $uv$, $(v,u)$ between vertices
$u,v$ have been set to forbidden, we set the remaining edge to permanent.  If
an edge is set to permanent and forbidden simultaneously or, equivalently, if
all three edges $(u,v)$, $uv$, $(v,u)$ are set to forbidden simultaneously,
then the instance has no solution with cost at most~$k$.  In case entries in
$M$ have been permanently set, we can extend these rules as follows: We
assume $\abs{N(u-v)} = \infty$ if both $M[t,u] = 1$ and $M[t,v] = 0$ are
permanent for some taxon~$t$; and $\abs{N(u+v)} = \infty$ if both $M[t,u] =
1$ and $M[t,v] = 1$ are permanent for some taxon~$t$.

On the other hand, we can use permanent edges in $G$ to derive information
about entries in~$M$: Keeping or setting some entry $M[t,u]$, will require us
to also change other entries in~$M$.  The next three rules follow
immediately:
\begin{rules}
\item \label{rule:induce-matrix-start} If $M[t,u] = 1$ is permanent and
  $(u,v)$ is permanent in $G$, then permanently set $M[t,v] = 1$.

\item If $M[t,v] = 0$ is permanent and $(u,v)$ is permanent in $G$, then
  permanently set $M[t,u] = 0$.

\item \label{rule:induce-matrix-end} If $M[t,u] = 1$ is permanent and $uv$ is
  permanent in $G$, then permanently set $M[t,v] = 0$.
\end{rules}
Again, if an entry $M[t,u]$ is permanently set to `0' and `1' simultaneously,
then the instance has no solution with cost at most~$k$.

Based on these observations, we can test in advance if the instance still
allows to permanently set an entry of the matrix to `0' or~`1'.  The
\emph{induced cost one} for entry $M[t,u]$, denoted $\ico(t,u)$, is the
number of vertices $v \in V$ such that $(u,v)$ is permanent and $M[t,v] = 0$,
plus the number of vertices $w \in V$ such that $uw$ is permanent and $M[t,w]
= 1$.  Similarly, we define the \emph{induced cost zero} for entry $M[t,v]$,
denoted $\icz(t,v)$, as the number of vertices $u \in V$ such that $(u,v)$ is
permanent and $M[t,u] = 1$.  We also take into account if the entry $M[t,v]$
is currently set to `0', `1', or~`?'.  To this end, we define $\ico_*(t,u) :=
\ico(t,u) +1$ if $M[t,u] = 0$, and $\ico_*(t,u) := \ico(t,u)$ otherwise.
Similarly, we define $\icz_*(t,v) := \icz(t,v) +1$ if $M[t,v] = 1$, and
$\icz_*(t,v) := \icz(t,v)$ otherwise.
\begin{rules}
\item \label{rule:induced-cost-start} If $\ico_*(t,u) > k$ then permanently
  set $M[t,u] = 0$.

\item \label{rule:induced-cost-end} If $\icz_*(t,v) > k$ then permanently
  set $M[t,v] = 1$.
\end{rules}

We can do the inverse reasoning of Rules
\ref{rule:induce-matrix-start}--\ref{rule:induce-matrix-end} and reach:
\begin{rules}
\item \label{rule:inverse-matrix-start} If $M[t,u] = 1$ is permanent and
  $M[t,v] = 0$ is permanent then set $(u,v)$ to forbidden.

\item \label{rule:inverse-matrix-end} If $M[t,u] = 1$ is permanent and
  $M[t,v] = 1$ is permanent then set $uv$ to forbidden.
\end{rules}

Finally, we can use the fact that the inclusion graph must be tree-ish:
\begin{rules}
\item \label{rule:treeish-start} If $(u,v)$ is permanent and $(v,w)$ is
  permanent then set $(u,w)$ to permanent.

\item If $(u,v)$ is permanent but $(u,w)$ is forbidden then set $(v,w)$ to
  forbidden.

\item If $(v,w)$ is permanent but $(u,w)$ is forbidden then set $(u,v)$ to
  forbidden.

\item If $(u,v)$ is permanent and $vw$ is permanent then set $uw$ to
  permanent.

\item If $(u,v)$ is permanent but $uw$ is forbidden then set $vw$ to
  forbidden.

\item \label{rule:treeish-end} If $vw$ is permanent but $uw$ is forbidden
  then set $(u,v)$ to forbidden.
\end{rules}

Finally, we can get rid of characters exhibited by a single taxon:

\begin{rules}
\item \label{rule:end} If a column in $M$ contains at most one `1' entry,
  then remove this column.
\end{rules}

Given an instance of MFST, we apply the above data reduction rules until the
conditions of none of the rules are met.  Whenever we change an entry of the
matrix $M$ by the above rules, we can lower our parameter $k$ by one which,
in turn, may allow us to apply other rules.  Still, the complete data
reduction requires only cubic time:

\begin{theorem}
  Rules \ref{rule:start}--\ref{rule:end} are correct, and can be carried out
  to completion in $O((m+n) m^2)$ time.
\end{theorem}

\begin{proof}
From the reasoning above, it is quite obvious that all rules are correct.
So, we focus on the running time of the data reduction.

Given an instance $M$ of the MFST problem, we first compute the inclusion
graph in time $O(m^2 n)$.  Note that in the matrix $M$, at most $O(mn)$
entries can be permanently set to `0' or `1' during the course of the data
reduction.  Similarly, at most $O(m^2)$ edges can be set to forbidden or
permanent in the inclusion graph.  Whenever we permanently flip an entry
in~$M$, we lower our cost bound $k$ by one.

Initially, we compute $q(u-v) := \abs{N(u-v)}$ and $q(u+v) := \abs{N(u+v)}$
for all $u,v$ in time $O(m^2 n)$.  Now, we can test Rules
\ref{rule:param-edges-start}--\ref{rule:param-edges-end} in constant times
for each pair $u,v$.  Similarly, we compute $\ico_*(t,v)$ and $\icz_*(t,v)$
for all $v,t$ in time $O(m^2 n)$, what allows us to test Rules
\ref{rule:induced-cost-start}--\ref{rule:induced-cost-end} in constant time
for each pair $t,v$.  During the course of our data reduction, the parameter
$k$ will change, so we have to efficiently find those pairs $u,v$ or $t,v$
that allow to use one of these rules.  For each value $0,\dots,k$ as well as
all values $> k$ we use an individual bin, and we use double-linked lists to
access those pairs that allow application of the above rules.  Updating
$q(u-v)$, $q(u+v)$, $\ico_*(t,v)$, and $\icz_*(t,v)$ can still be performed
in constant time.  So, in constant time we can find a pair $u,v$ or $t,v$ to
apply a reduction rule, or decide that no such pair exists.

All other rules are only applied if a matrix entry is permanently set or
flipped, or if an edge in the inclusion graph is set to forbidden or
permanent.  For each rule, we now analyze under what circumstances it can be
applied, an what time is required to apply the rule.

For Rules \ref{rule:param-edges-start}--\ref{rule:param-edges-end} we have to
update $q(u-v)$ and $q(u+v)$ every time an entry in the matrix is flipped.
Assume that $M[t,u]$ is the matrix entry being flipped, then we check for all
$v \ne u$, whether $q(u-v)$, $q(v-u)$, or $q(u+v)$ must be updated.  In this
case, these values are increased or decreased by one, depending on the entry
$M[t,v]$.  In total, a flip in the matrix $M$ requires $O(m)$ time to update
all $q(u-v)$ and $q(u+v)$.

Rules \ref{rule:induce-matrix-start}--\ref{rule:induce-matrix-end} must be
applied if either a matrix entry is permanently set, or if an edge is set to
forbidden or permanent.  Regarding Rule~\ref{rule:induce-matrix-start},
assume that $M[t,u]$ is permanently set to `1'.  In this case, we have to
test for all $v \ne u$ if $(u,v)$ is permanent in the inclusion graphs, what
can be done in time $O(m)$.  Now, assume that some edge $(u,v)$ is set to
permanent.  Then, we have to check all taxa $t$ if $M[t,u] = 1$ is permanent,
what can be done in time $O(n)$.  A similar reasoning applies for the other
two rules.

Rules \ref{rule:induced-cost-start}--\ref{rule:induced-cost-end} require us
to update $\ico_*(t,v)$ and $\icz_*(t,u)$ whenever either a matrix entry is
flipped, or an edge is set to permanent.  Regarding $\ico_*(\cdot)$, assume
that entry $M[t,v]$ has been flipped to~`0'.  Then, for all $u \ne v$ such
that $(u,v)$ is permanent, we increase $\ico_*(t,u)$ by one.  Similarly, for
all $w \ne v$ such that $vw$ is permanent, we decrease $\ico_*(t,w)$ by one.
If $M[t,v]$ has been flipped to~`1' then we do the same, exchanging increase
and decrease.  This can be carried out in time $O(m)$.  If an edge is set to
permanent, we can update all affected entries in time $O(n)$.  A similar
reasoning applies for the computation of $\icz_*(t,u)$.

For Rules \ref{rule:inverse-matrix-start}--\ref{rule:inverse-matrix-end} we
have to update edges in case an entry $M[t,v]$ is flipped: Then, we have to
consider all entries $M[t,u]$ for $u \ne v$ what can be done in time $O(m)$.

Rules \ref{rule:treeish-start}--\ref{rule:treeish-end} update edges in case
some edge between $u$ and $v$ is set to forbidden or permanent: Then, we have
to consider all vertices $w \ne u,v$ what requires $O(m)$ time.

Applying the above rules, may result in more than one ``update operations''
to be carried out.  For that, we can keep all such update operations on a
stack, and carry out the next update operation only after we have finished
the current one.

We conclude that permanently setting an entry of the matrix requires $O(m)$
time for checking all of the rules.  Since we can permanently set at most
$O(mn)$ entries, this requires $O(m^2 n)$ time in total.  Similarly,
setting an edge to permanent, requires $O(m+n)$ time for checking our rules.
Since there are $O(m^2)$ edges the total running time becomes $O((m+n) m^2)$.
This results in a running time of $O((m+n) m^2)$ for the full data reduction.
\qed
\end{proof}

If we reach a conflict in our data reduction, such as permanently setting
some $M[t,v]$ to $0$ and $1$ at the same time, then we infer that there
exists no solution of the instance of cost at most~$k$.


\section{Upper and lower bounds} \label{sec:bounds}

We will now describe a lower bound for the \textsc{Minimum Flip Supertree}
problem, which we will use to derive improved versions of Rules
\ref{rule:param-edges-start}--\ref{rule:param-edges-end} and
\ref{rule:induced-cost-start}--\ref{rule:induced-cost-end}.  A \emph{local
  conflict} consists of two characters $u,v \in \{1,\dots,m\}$ and three taxa
$t_1, t_2, t_3 \in \{1,\dots,n\}$ such that $M[t_1,u] = M[t_2,u] = M[t_2,v] =
M[t_3,v] = 1$ but $M[t_3,u] = M[t_1,v] = 0$.  In the \textsc{Minimum Flip
  Consensus Tree} setting, $M$ admits a perfect phylogeny if and only if $M$
does not contain a local conflict~\cite{chen06minimum-flip}.  For MFST, we
can only reason that if $M$ contains a local conflict, then it does not admit
a perfect phylogeny.

We now use local conflicts to compute a lower bound for the costs of an
instance~$M$: We say that two local conflicts are \emph{edge-disjoint} if the
local conflicts do not contain a common tuple $(v,t)$.  The term
``edge-disjoint'' stems from visualizing the matrix $M$ as a bipartite
graph~\cite{chen06minimum-flip}, as noted in Sec.~\ref{sec:preliminaries}.
Let $\cC$ be a set of edge-disjoint local conflicts in $M$.  Now, for every
element in $\cC$ we have to make at least one modification to the matrix $M$
to remove the local conflict, so $\abs{\cC}$ is a lower bound to the cost of
an optimal solution.  Unfortunately, it is not obvious how to efficiently
find a set $\cC$ of maximal cardinality: For example, the obvious
transformation to a graph leads to the NP-hard \textsc{Maximum Independent
  Set} problem.  In case columns of $M$ have been weighted, we can follow a
\emph{greedy} strategy, choosing a local conflict that maximizes the cost of
the current step.  We can also weight each local conflict by the (inverse)
number of other local conflicts it has at least one common edge with.
Finally, we can restart the algorithm several times, choosing a random local
conflict in each step, and maximize over these bounds.  In theory, we can
compute another lower bound by solving the relaxation of the Integer Linear
Program presented in~\cite{chimani10exact}, but this is usually too slow in
practice.

Testing all characters $u,v$ and taxa $t_1, t_2, t_3$ is prohibitive in
application, as this requires $O(m^2 n^3)$ time.  But we can use sets
$N(u-v)$ and $N(u+v)$ for this purpose: Two characters are in conflict if all
sets $N(u-v)$, $N(u+v)$, and $N(v-u)$ contain at least one element.  We now
describe an improved algorithm for the greedy strategy and its variants
discussed above.  Initially, we compute the cardinality $q(u-v)$ and $q(u+v)$
in $O(m^2 n)$ time, storing $q$ requires $O(m^2)$ space.  We start with a set
$\cL = \{\{u,v\} \;:\; 1 \le u < v \le m\}$ of character pairs that are
potentially in conflict.  We then select a pair $\{u,v\}$ from $\cL$, either
randomly or by some other criterion.  If $u,v$ are no longer in conflict, we
remove $\{u,v\}$ from~$\cL$.  Otherwise, we choose a certain local conflict
$u,v,t_1,t_2,t_3$ to be part of our set $\cC$: We then update, for each tuple
$(w,t)$ of the local conflict, $w \in \{u,v\}$ and $t \in \{t_1,t_2,t_3\}$,
all cardinalities $q(w-w')$, $q(w+w')$, and $q(w'-w)$ for all characters $w'$
in time $O(mn)$.  Let $k_{\text{opt}}$ be the cost of an optimal solution,
then $\abs{\cC} \le k_{\text{opt}} \le mn$.  All updates require $O(\abs{\cC}
mn)$ time and, hence, $O(k_{\text{opt}} mn)$ time.  The whole procedure
requires only $O(k_{\text{opt}} mn + m^2 n)$ time in total.  In practice, we
can speed up calculations by initializing $\cL$ with those tuples $\{u,v\}$
that are initially in conflict.  Also, note that our data reduction requires
us to maintain cardinalities $q$ before we enter the computation of a lower
bound.

\medskip

We now use a trick introduced in~\cite{boecker11exact} to lift a local
reduction rule to a global version:
Rules \ref{rule:param-edges-start}--\ref{rule:param-edges-end} are
\emph{local}, in the sense that these rules only take into account entries
$M[t,u]$ and $M[t,v]$ for all taxa~$t$.  Similarly, computing $\ico_*(t,v)$
and $\icz_*(t,v)$ will consider entries $M[t,w]$ for all characters~$w$, but
ignore the rest of the matrix.  As all other rows or columns of $M$ have to
be cleaned of local conflicts at a later stage, it makes sense to estimate
the cost for doing so using a lower bound.

Let $lb_M(t)$ be any lower bound where, during the calculation of this bound,
the row of $M$ corresponding to taxon $t$ is not taken into account.
Similarly, we write $lb_M(u,v)$ for two ignored character columns $u,v$.
Now, we can write improved versions of Rules
\ref{rule:param-edges-start}--\ref{rule:param-edges-end} and
\ref{rule:induced-cost-start}--\ref{rule:induced-cost-end}:
\begin{rules}
\item \label{rule:lb-start} If $\abs{N(u-v)} + lb_M(u,v) > k$ then set
  $(u,v)$ to forbidden.

\item If $\abs{N(u+v)} + lb_M(u,v) > k$ then set $uv$ to forbidden.

\item \label{rule:lb-induced-cost-start} If $\ico_*(t,v) + lb_M(t) > k$ then
  permanently set $M[t,v] = 0$.

\item \label{rule:lb-induced-cost-end} \label{rule:lb-end} If $\icz_*(t,u) +
  lb_M(t) > k$ then permanently set $M[t,u] = 1$.
\end{rules}
The correctness of these rules follows immediately.  Unfortunately, we have
to compute an individual lower bound $lb_M(u,v)$ for every pair $u,v$ and
$lb_M(t)$ for every~$t$.  To further speed up calculations, we can initially
compute a lower bound $lb_M$ of the complete instance, and calculate
$lb_M(u,v)$ only for those pairs where $\abs{N(u-v)} + lb_M > k$ or
$\abs{N(u+v)} + lb_M > k$ holds.  Note that there may exist rare cases where
our lower bond computations are not monotonous, so that $lb_M(u,v) > lb_M$,
and we will miss a rule that could have been applied.  We expect this to be
negligible in practice.  A similar reasoning applies for $lb_M(t)$.

\medskip

The above rules still depend on parameter~$k$.  To reach a
parameter-independent data reduction, we have to choose an appropriate~$k$:
To this end, note that the cost of any heuristic solution to an instance, are
always an upper bound to the cost of an optimal solution.  So, we can choose
any heuristic to compute an appropriate~$k$, and then apply our parameter
dependent data reduction, using a lower bound for Rules
\ref{rule:lb-start}--\ref{rule:lb-end}.  It must be understood that for
practically all real-world instances, the cost $k$ computed by any heuristic
will be too large to directly apply Rules
\ref{rule:param-edges-start}--\ref{rule:param-edges-end} and
\ref{rule:induced-cost-start}--\ref{rule:induced-cost-end}.  Only through our
algorithm engineering technique of using lower bounds, we can successfully
start our data reduction.  Rules \ref{rule:lb-induced-cost-start} and
\ref{rule:lb-induced-cost-end} will sometimes allow us to lower the cost $k$
and, hence, the complexity of the remaining instance.

Chen \emph{et~al.}~\cite{chen06improved-heuristics} have introduced an
involved heuristic for the problem that, much like heuristics for the Maximum
Parsimony problem, is based on exploring tree space via branch swapping.
This heuristic is rather time-consuming and can require minutes or even hours
of running time, but its results are of excellent
quality~\cite{chimani10exact}.  Another upper bound can be computed by
running the ILP from~\cite{chimani10exact} for some time, and stop after a
fixed time before upper and lower bound of the instance coincide.


\section{Experiments} \label{sec:experiments}

We implemented all evaluated algorithms in Java.  Computations were performed
on an AMD Opteron-275 $2.2$~GHz with $6$~GB of memory running Solaris~10.


We now evaluate the parameter-independent data reduction.  As indicated in
the introduction, we can use our reduction as a preprocessing step, and solve
the reduced instance with any exact, approximation, or heuristic algorithm.
To evaluate the performance of our data reduction, we use different measures:
\begin{itemize}
\item We calculate the ratio of \emph{fixed} entries, `0/1' entries that
  are not flipped by the data reduction but set to permanent, relative to the
  number of `0/1' entries in the input matrix.

\item Similarly, we calculate the ratio of \emph{flipped} entries, which are
  always permanent.

\item We calculate the ratio of \emph{resolved} entries, `?' entries that are
  permanently set to `0/1', relative to the number of `?' entries in the
  input matrix.

\item We calculate the ratio of \emph{permanent} entries in the matrix,
  relative to the size $mn$ of the matrix.

\item Next, we count how many edges in the inclusion graph have been set to
  permanent, and compare this to the $\binom{m}{2}$ possible edges.
  
\item For all pairs $u,v$ where no permanent edge exists, we count the number
  of forbidden edges, and compare this to the $3\binom{m}{2}$ possible
  forbidden edges.

\item Finally, we calculate the number of flips executed by the data
  reduction, and compare it to the number of flips required to solve the
  instance.  This reduces the cost and, hence, the complexity of the
  resulting instance.
\end{itemize}

For our evaluation, we use instances generated by Eulenstein
\emph{et~al.}~\cite{eulenstein04performance}, see there for details.  These
simulated datasets are very similar to a regular phylogenetic supertree
study, yet for each dataset we know the \emph{true} model tree behind the
data.  Unfortunately, running times of our data reduction are currently
prohibitive for larger instances as well as instances with a large fraction
of `?'.  To this end, we concentrate on matrices containing $25\,\%$
`?'-entries, generated from $s = 4,6,8$ input trees.  The number of taxa $n$
is either 48 or 96.  These matrices contain about $m \approx (pn-2) s$
columns, where $1-p$ is the ratio of `?'-entries.
For each parameter combination, we choose the 100 instances named ``random'',
for which deleted taxa were randomly chosen.  We use the heuristic solutions
from~\cite{chen06improved-heuristics} as upper bounds for our
parameter-independent data reduction, and the randomized lower bound from
Sec.~\ref{sec:bounds} with 100 repetitions.


\begin{table}[tb]
\begin{center}\small
\begin{tabular}{@{}l | rrr | r @{}}
  Taxa $n$ & 48 & 48 & 48 & 96 \\

  Input trees $s$ & 4 & 6 & 8 & 4 \\

\hline

  Fixed entries & 36.13 & 30.43 & 24.64 & 7.46 \\

  Flipped entries & 0.04 & 0.04 & 0.04 & 0.002 \\

  Resolved entries & 31.90 & 28.69 & 23.69 & 6.23 \\

  Permanent entries & 35.12 & 30.03 & 24.43 & 7.26 \\ 

  Permanent edges & 29.28 & 25.46 & 20.20 & 4.92 \\

  Forbidden edges & 28.22 & 23.63 & 19.03 & 7.68 \\[0.3ex]

  Number flips & 2.02 & 3.07 & 3.51 & 0.51 \\

  No.~flips relative & 12.26 & 9.88 & 8.78 & 1.11 \\[0.3ex]

  Running time (h:min) & 0:50 & 5:40 & 22:51 & 24:08
\end{tabular}
\end{center}
\caption{Results of the data reduction for $25\,\%$ taxa deletion.  Averages
  over 100 instances.  All numbers except ``number flips'' and ``running
  time'' in percent.}
  \label{tab:reduction-results}
\end{table}


One can see that reduction ratios deteriorate for both increasing number of
input trees, and increasing number of taxa.  We expect this to be even more
so for higher ratios of taxa deletion.  Currently, the limiting factor are
the high running times of the data reduction.  On the other hand, we
observe that the data reduction truly does reduce the instances.  This is a
clear indication that with an improved implementation, algorithm engineering,
new data reduction rules, and an improved lower bound, we may indeed simplify
MFST instances in polynomial time.


\section{Conclusion}

We have presented a set of data reduction rules, that allow us preprocess
instances of the \textsc{Minimum Flip Supertree} problem, and also of the
``simpler'' \textsc{Minimum Flip Consensus Tree} problem.  Our data reduction
can be applied in polynomial running time.  Different
from~\cite{komusiewicz08cubic}, our reduction allows us to draw conclusions
about certain entries in the input matrix.  This is highly desirable, as
flipping entries during preprocessing means that we are reducing the cost of
the resulting instance: Chimani \emph{et al.}~\cite{chimani10exact} found
that ILP running times are strongly correlated with the optimal number of
flips.

Our method allows us to draw conclusions about MFST instances, guaranteeing
both polynomial running time and optimality of the solution.  On the
practical side, the output of our method can be subsequently processed with
any method, including fast heuristics.  Unfortunately, our reduction is
currently not suited for real-world application, as running times are
prohibitive and reduction results are minor.  Still, we think that this an
important first step towards a data reduction that is applicable in practice.
An improved data reduction may be ultimately combined with heuristics to
obtain a supertree method that is both fast and accurate in practice.

Note that the inclusion graph can also be used as an algorithm engineering
technique for the MFCT search tree algorithm from~\cite{boecker08improved}.
We conjecture that these techniques will make the search tree algorithm much
faster in practice.  Unfortunately, it appears this cannot be used to improve
upon the worst-case running time of the algorithm.

We conjecture that our data reduction from Sec.~\ref{sec:data-reduction} can
be used as part of a problem kernel for the MFCT problem.  From the
theoretical side, it is an interesting open question if this allows us to
find a better than cubic kernel~\cite{komusiewicz08cubic}.  Finding a kernel
for the MFST problem, on the other hand, is related to the open question
whether MFST is parameterized tractable.

\paragraph{Acknowledgment.}

Implementation by Konstantin Riege and Andreas Dix.

\bibliographystyle{abbrv}
\bibliography{bibtex/group-literature}

\end{document}